\documentclass[useAMS,usenatbib]{mn2e}

\usepackage{times}
\usepackage{graphicx}
\graphicspath{{./fig/}}
\usepackage{bm}
\usepackage{url}

\title
[Random localized expansion wave turbulence]
{Turbulence from localized random expansion waves}

\author[Antony J.\ Mee and Axel Brandenburg]%
{Antony J.\ Mee$^1$ and Axel Brandenburg$^2$\\
$^1$School of Mathematics and Statistics, University of Newcastle,
  Newcastle upon Tyne, NE1 7RU, UK\\
$^2$NORDITA, Blegdamsvej 17, DK-2100 Copenhagen \O, Denmark}

\begin{document}


\newcommand{\EQ}{\begin{equation}}
\newcommand{\EN}{\end{equation}}
\newcommand{\EQA}{\begin{eqnarray}}
\newcommand{\ENA}{\end{eqnarray}}
\newcommand{\eq}[1]{(\ref{#1})}
\newcommand{\EEq}[1]{Equation~(\ref{#1})}
\newcommand{\Eq}[1]{Eq.~(\ref{#1})}
\newcommand{\Eqs}[2]{Eqs~(\ref{#1}) and~(\ref{#2})}
\newcommand{\Eqss}[2]{Eqs~(\ref{#1})--(\ref{#2})}
\newcommand{\eqs}[2]{(\ref{#1}) and~(\ref{#2})}
\newcommand{\App}[1]{Appendix~\ref{#1}}
\newcommand{\Sec}[1]{Sect.~\ref{#1}}
\newcommand{\Secs}[2]{Sects\ref{#1} and \ref{#2}}
\newcommand{\Fig}[1]{Fig.~\ref{#1}}
\newcommand{\FFig}[1]{Figure~\ref{#1}}
\newcommand{\Tab}[1]{Table~\ref{#1}}
\newcommand{\Figs}[2]{Figs~\ref{#1} and \ref{#2}}
\newcommand{\Tabs}[2]{Tables~\ref{#1} and \ref{#2}}
\newcommand{\bra}[1]{\langle #1\rangle}
\newcommand{\bbra}[1]{\left\langle #1\right\rangle}
\newcommand{\mean}[1]{\overline #1}
\newcommand{\meanemf}{\overline{\mbox{\boldmath ${\cal E}$}}{}}{}
\newcommand{\meanemfs}{\overline{\cal E} {}}
\newcommand{\meanAA}{\overline{\bm{A}}}
\newcommand{\meanBB}{\overline{\bm{B}}}
\newcommand{\meanJJ}{\overline{\bm{J}}}
\newcommand{\meanUU}{\overline{\bm{U}}}
\newcommand{\meanWW}{\overline{\bm{W}}}
\newcommand{\meanFF}{\overline{{\cal{\bm{F}}}}}
\newcommand{\meanuu}{\overline{\mbox{\boldmath $u$}}{}}{}
\newcommand{\meanoo}{\overline{\mbox{\boldmath $\omega$}}{}}{}
\newcommand{\meanEE}{\overline{\mbox{\boldmath ${\cal E}$}}{}}{}
\newcommand{\meanuxB}{\overline{\mbox{\boldmath $\delta u\times \delta B$}}{}}{}
\newcommand{\meanJB}{\overline{\mbox{\boldmath $J\cdot B$}}{}}{}
\newcommand{\meanAB}{\overline{\mbox{\boldmath $A\cdot B$}}{}}{}
\newcommand{\meanjb}{\overline{\mbox{\boldmath $j\cdot b$}}{}}{}
\newcommand{\meanA}{\overline{A}}
\newcommand{\meanB}{\overline{B}}
\newcommand{\meanC}{\overline{C}}
\newcommand{\meanU}{\overline{U}}
\newcommand{\meanJ}{\overline{J}}
\newcommand{\meanS}{\overline{S}}
\newcommand{\meanF}{\overline{\cal F}}
%
%
\newcommand{\teps}{\tilde{\epsilon} {}}
\newcommand{\zh}{\hat{z}}
%
%
\newcommand{\eee}{\hat{\mbox{\boldmath $e$}} {}}
\newcommand{\nnn}{\hat{\mbox{\boldmath $n$}} {}}
\newcommand{\vvv}{\hat{\mbox{\boldmath $v$}} {}}
\newcommand{\rr}{\hat{\mbox{\boldmath $r$}} {}}
\newcommand{\xxx}{\hat{\bm x}}
\newcommand{\yyy}{\hat{\bm y}}
\newcommand{\zz}{\hat{\bm z}}
\newcommand{\pphi}{\hat{\bm\phi}}
\newcommand{\ttt}{\hat{\mbox{\boldmath $\theta$}} {}}
\newcommand{\OOO}{\hat{\mbox{\boldmath $\Omega$}} {}}
\newcommand{\ooo}{\hat{\mbox{\boldmath $\omega$}} {}}
\newcommand{\BBBB}{\hat{\mbox{\boldmath $B$}} {}}
\newcommand{\Bhat}{\hat{B}}
\newcommand{\BBhat}{\hat{\bm{B}}}
%
%
\newcommand{\gggg}{\mbox{\boldmath $g$} {}}
\newcommand{\ddd}{\mbox{\boldmath $d$} {}}
\newcommand{\rrr}{\mbox{\boldmath $r$} {}}
\newcommand{\yy}{\mbox{\boldmath $y$} {}}
\newcommand{\zzz}{\mbox{\boldmath $z$} {}}
\newcommand{\vv}{\mbox{\boldmath $v$} {}}
\newcommand{\ww}{\mbox{\boldmath $w$} {}}
\newcommand{\mm}{\mbox{\boldmath $m$} {}}
\newcommand{\PP}{\mbox{\boldmath $P$} {}}
\newcommand{\bp}{\mbox{\boldmath $p$} {}}
\newcommand{\pp}{\mbox{\boldmath $p$} {}}
\newcommand{\II}{\mbox{\boldmath $I$} {}}
\newcommand{\qq}{{\bm{q}}}
\newcommand{\xx}{{\bm{x}}}
\newcommand{\UU}{{\bm{U}}}
\newcommand{\WW}{{\bm{W}}}
\newcommand{\QQ}{{\bm{Q}}}
\newcommand{\uu}{{\bm{u}}}
\newcommand{\BB}{{\bm{B}}}
\newcommand{\HH}{{\bm{H}}}
\newcommand{\CC}{{\bm{C}}}
\newcommand{\JJ}{{\bm{J}}}
\newcommand{\jj}{{\bm{j}}}
\newcommand{\AAA}{{\bm{A}}}
\newcommand{\aaaa}{{\bm{a}}}
\newcommand{\bb}{{\bm{b}}}
\newcommand{\cc}{{\bm{c}}}
\newcommand{\ee}{{\bm{e}}}
\newcommand{\nn}{\mbox{\boldmath $n$} {}}
\newcommand{\ff}{\mbox{\boldmath $f$} {}}
\newcommand{\hh}{\mbox{\boldmath $h$} {}}
\newcommand{\EE}{{\bm{E}}}
\newcommand{\FF}{{\bm{F}}}
\newcommand{\FFF}{{\bm{{\cal F}}}}
\newcommand{\KK}{{\bm{K}}}
\newcommand{\kk}{{\bm{k}}}
\newcommand{\TT}{\mbox{\boldmath $T$} {}}
\newcommand{\MM}{\mbox{\boldmath $M$} {}}
\newcommand{\GG}{{\bm{G}}}
\newcommand{\SSS}{{\bm{S}}}
\newcommand{\grav}{\mbox{\boldmath $g$} {}}
\newcommand{\nab}{\mbox{\boldmath $\nabla$} {}}
\newcommand{\OO}{\mbox{\boldmath $\Omega$} {}}
\newcommand{\oo}{\mbox{\boldmath $\omega$} {}}
\newcommand{\ttau}{\mbox{\boldmath $\tau$} {}}
\newcommand{\LL}{\mbox{\boldmath $\Lambda$} {}}
\newcommand{\llambda}{\mbox{\boldmath $\lambda$} {}}
\newcommand{\pomega}{\mbox{\boldmath $\varpi$} {}}
%
%
\newcommand{\SSSS}{\mbox{\boldmath $\sf S$} {}}
\newcommand{\RRRR}{\mbox{\boldmath $\sf R$} {}}
\newcommand{\LLLL}{\mbox{\boldmath $\sf L$} {}}
\newcommand{\PPPP}{\mbox{\boldmath ${\sf P}$} {}}
\newcommand{\MMMM}{\mbox{\boldmath ${\sf M}$} {}}
\newcommand{\AAAA}{\mbox{\boldmath ${\cal A}$} {}}
\newcommand{\BBB}{\mbox{\boldmath ${\cal B}$} {}}
\newcommand{\emf}{\mbox{\boldmath ${\cal E}$} {}}
\newcommand{\GGG}{\mbox{\boldmath ${\cal G}$} {}}
\newcommand{\HHH}{\mbox{\boldmath ${\cal H}$} {}}
\newcommand{\QQQ}{\mbox{\boldmath ${\cal Q}$} {}}
\newcommand{\GGGG}{{\bf G} {}}
%
%
\newcommand{\ii}{{\rm i}}
\newcommand{\grad}{{\rm grad} \, {}}
\newcommand{\curl}{{\rm curl} \, {}}
\newcommand{\dive}{{\rm div}  \, {}}
\newcommand{\Dive}{{\rm Div}  \, {}}
\newcommand{\sgn}{{\rm sgn}  \, {}}
\newcommand{\DD}{{\rm D} {}}
\newcommand{\DDD}{{\cal D} {}}
\newcommand{\dd}{{\rm d} {}}
\newcommand{\const}{{\rm const}  {}}
\newcommand{\CR}{{\rm CR}}
\def\degr{\hbox{$^\circ$}}
\def\la{\mathrel{\mathchoice {\vcenter{\offinterlineskip\halign{\hfil
$\displaystyle##$\hfil\cr<\cr\sim\cr}}}
{\vcenter{\offinterlineskip\halign{\hfil$\textstyle##$\hfil\cr<\cr\sim\cr}}}
{\vcenter{\offinterlineskip\halign{\hfil$\scriptstyle##$\hfil\cr<\cr\sim\cr}}}
{\vcenter{\offinterlineskip\halign{\hfil$\scriptscriptstyle##$\hfil\cr<\cr\sim\cr}}}}}
\def\ga{\mathrel{\mathchoice {\vcenter{\offinterlineskip\halign{\hfil
$\displaystyle##$\hfil\cr>\cr\sim\cr}}}
{\vcenter{\offinterlineskip\halign{\hfil$\textstyle##$\hfil\cr>\cr\sim\cr}}}
{\vcenter{\offinterlineskip\halign{\hfil$\scriptstyle##$\hfil\cr>\cr\sim\cr}}}
{\vcenter{\offinterlineskip\halign{\hfil$\scriptscriptstyle##$\hfil\cr>\cr\sim\cr}}}}}
%
%
\def\Ta{\mbox{\rm Ta}}
\def\Ra{\mbox{\rm Ra}}
\def\Ma{\mbox{\rm Ma}}
\def\Co{\mbox{\rm Co}}
\def\Roo{\mbox{\rm Ro}^{-1}}
\def\Rooo{\mbox{\rm Ro}^{-2}}
\def\Pra{\mbox{\rm Pr}}
\def\Pran{\mbox{\rm Pr}}
\def\Pm{\mbox{\rm Pr}_M}
\def\Rm{\mbox{\rm Re}_M}
\def\Rey{\mbox{\rm Re}}
\def\Pe{\mbox{\rm Pe}}
\newcommand{\ea}{{\rm et al.\ }}
\newcommand{\eaa}{{\rm et al.\ }}
\def\half{{\textstyle{1\over2}}}
\def\threehalf{{\textstyle{3\over2}}}
\def\onethird{{\textstyle{1\over3}}}
\def\onesixth{{\textstyle{1\over6}}}
\def\twothird{{\textstyle{2\over3}}}
\def\fourthird{{\textstyle{4\over3}}}
\def\quarter{{\textstyle{1\over4}}}
\newcommand{\W}{\,{\rm W}}
\newcommand{\V}{\,{\rm V}}
\newcommand{\kV}{\,{\rm kV}}
\newcommand{\T}{\,{\rm T}}
\newcommand{\G}{\,{\rm G}}
\newcommand{\Hz}{\,{\rm Hz}}
\newcommand{\nHz}{\,{\rm nHz}}
\newcommand{\kHz}{\,{\rm kHz}}
\newcommand{\kG}{\,{\rm kG}}
\newcommand{\mkG}{\,\mu{\rm G}}
\newcommand{\K}{\,{\rm K}}
\newcommand{\g}{\,{\rm g}}
\newcommand{\s}{\,{\rm s}}
\newcommand{\ms}{\,{\rm ms}}
\newcommand{\mpers}{\,{\rm m/s}}
\newcommand{\ks}{\,{\rm ks}}
\newcommand{\cm}{\,{\rm cm}}
\newcommand{\m}{\,{\rm m}}
\newcommand{\km}{\,{\rm km}}
\newcommand{\kms}{\,{\rm km/s}}
\newcommand{\kg}{\,{\rm kg}}
\newcommand{\ug}{\,\mu{\rm g}}
\newcommand{\kW}{\,{\rm kW}}
\newcommand{\MW}{\,{\rm MW}}
\newcommand{\Mm}{\,{\rm Mm}}
\newcommand{\Mx}{\,{\rm Mx}}
\newcommand{\pc}{\,{\rm pc}}
\newcommand{\kpc}{\,{\rm kpc}}
\newcommand{\yr}{\,{\rm yr}}
\newcommand{\Myr}{\,{\rm Myr}}
\newcommand{\Gyr}{\,{\rm Gyr}}
\newcommand{\erg}{\,{\rm erg}}
\newcommand{\mol}{\,{\rm mol}}
\newcommand{\dyn}{\,{\rm dyn}}
\newcommand{\J}{\,{\rm J}}
\newcommand{\RM}{\,{\rm RM}}
\newcommand{\EM}{\,{\rm EM}}
\newcommand{\AU}{\,{\rm AU}}
\newcommand{\A}{\,{\rm A}}
%
%
\newcommand{\yastroph}[2]{ #1, astro-ph/#2}
\newcommand{\ycsf}[3]{ #1, {Chaos, Solitons \& Fractals,} {#2}, #3}
\newcommand{\yepl}[3]{ #1, {Europhys. Lett.,} {#2}, #3}
\newcommand{\yaj}[3]{ #1, {AJ,} {#2}, #3}
\newcommand{\yjgr}[3]{ #1, {JGR,} {#2}, #3}
\newcommand{\ysol}[3]{ #1, {Sol. Phys.,} {#2}, #3}
\newcommand{\yapj}[3]{ #1, {ApJ,} {#2}, #3}
\newcommand{\ypasp}[3]{ #1, {PASP,} {#2}, #3}
\newcommand{\yapjl}[3]{ #1, {ApJ,} {#2}, #3}
\newcommand{\yapjs}[3]{ #1, {ApJS,} {#2}, #3}
\newcommand{\yan}[3]{ #1, {AN,} {#2}, #3}
\newcommand{\yzfa}[3]{ #1, {Z.\ f.\ Ap.,} {#2}, #3}
\newcommand{\ymhdn}[3]{ #1, {Magnetohydrodyn.} {#2}, #3}
\newcommand{\yana}[3]{ #1, {A\&A,} {#2}, #3}
\newcommand{\yanas}[3]{ #1, {A\&AS,} {#2}, #3}
\newcommand{\yanar}[3]{ #1, {A\&AR,} {#2}, #3}
\newcommand{\yass}[3]{ #1, {Ap\&SS,} {#2}, #3}
\newcommand{\ygafd}[3]{ #1, {Geophys. Astrophys. Fluid Dyn.,} {#2}, #3}
\newcommand{\ypasj}[3]{ #1, {Publ. Astron. Soc. Japan,} {#2}, #3}
\newcommand{\yjfm}[3]{ #1, {JFM,} {#2}, #3}
\newcommand{\ypf}[3]{ #1, {Phys. Fluids,} {#2}, #3}
\newcommand{\ypp}[3]{ #1, {Phys. Plasmas,} {#2}, #3}
\newcommand{\ysov}[3]{ #1, {Sov. Astron.,} {#2}, #3}
\newcommand{\ysovl}[3]{ #1, {Sov. Astron. Lett.,} {#2}, #3}
\newcommand{\yjetp}[3]{ #1, {Sov. Phys. JETP,} {#2}, #3}
\newcommand{\yphy}[3]{ #1, {Physica,} {#2}, #3}
\newcommand{\yannr}[3]{ #1, {ARA\&A,} {#2}, #3}
\newcommand{\yaraa}[3]{ #1, {ARA\&A,} {#2}, #3}
\newcommand{\yprs}[3]{ #1, {Proc. Roy. Soc. Lond.,} {#2}, #3}
\newcommand{\yprl}[3]{ #1, {Phys. Rev. Lett.,} {#2}, #3}
\newcommand{\yphl}[3]{ #1, {Phys. Lett.,} {#2}, #3}
\newcommand{\yptrs}[3]{ #1, {Phil. Trans. Roy. Soc.,} {#2}, #3}
\newcommand{\ymn}[3]{ #1, {MNRAS,} {#2}, #3}
\newcommand{\ynat}[3]{ #1, {Nat,} {#2}, #3}
\newcommand{\ysci}[3]{ #1, {Sci,} {#2}, #3}
\newcommand{\ysph}[3]{ #1, {Solar Phys.,} {#2}, #3}
\newcommand{\ypr}[3]{ #1, {Phys. Rev.,} {#2}, #3}
\newcommand{\ypre}[3]{ #1, {Phys. Rev. E,} {#2}, #3}
\newcommand{\spr}[2]{ ~#1~ {\em Phys. Rev. }{\bf #2} (submitted)}
\newcommand{\ppr}[2]{ ~#1~ {\em Phys. Rev. }{\bf #2} (in press)}
\newcommand{\ypnas}[3]{ #1, {Proc. Nat. Acad. Sci.,} {#2}, #3}
\newcommand{\yicarus}[3]{ #1, {Icarus,} {#2}, #3}
\newcommand{\yspd}[3]{ #1, {Sov. Phys. Dokl.,} {#2}, #3}
\newcommand{\yjcp}[3]{ #1, {J. Comput. Phys.,} {#2}, #3}
\newcommand{\yjour}[4]{ #1, {#2}, {#3}, #4}
\newcommand{\yprep}[2]{ #1, {\sf #2}}
\newcommand{\ybook}[3]{ #1, {#2} (#3)}
\newcommand{\yproc}[5]{ #1, in {#3}, ed. #4 (#5), #2}
\newcommand{\pproc}[4]{ #1, in {#2}, ed. #3 (#4), (in press)}
\newcommand{\ppp}[1]{ #1, {Phys. Plasmas,} (in press)}
\newcommand{\sapj}[1]{ #1, {ApJ,} (submitted)}
\newcommand{\sana}[1]{ #1, {A\&A,} (submitted)}
\newcommand{\san}[1]{ #1, {AN,} (submitted)}
\newcommand{\sprl}[1]{ #1, {PRL,} (submitted)}
\newcommand{\pprl}[1]{ #1, {PRL,} (in press)}
\newcommand{\sjfm}[1]{ #1, {JFM,} (submitted)}
\newcommand{\sgafd}[1]{ #1, {Geophys. Astrophys. Fluid Dyn.,} (submitted)}
\newcommand{\pgafd}[1]{ #1, {Geophys. Astrophys. Fluid Dyn.,} (in press)}
\newcommand{\tana}[1]{ #1, {A\&A,} (to be submitted)}
\newcommand{\smn}[1]{ #1, {MNRAS,} (submitted)}
\newcommand{\pmn}[1]{ #1, {MNRAS,} (in press)}
\newcommand{\papj}[1]{ #1, {ApJ,} (in press)}
\newcommand{\papjl}[1]{ #1, {ApJL,} (in press)}
\newcommand{\sapjl}[1]{ #1, {ApJL,} (submitted)}
\newcommand{\pana}[1]{ #1, {A\&A,} (in press)}
\newcommand{\pan}[1]{ #1, {AN,} (in press)}
\newcommand{\pjour}[2]{ #1, {#2,} (in press)}
\newcommand{\sjour}[2]{ #1, {#2,} (submitted)}

\date{\today,~ $ $Revision: 1.69 $ $}
\pagerange{\pageref{firstpage}--\pageref{lastpage}}
\pubyear{2006}

\maketitle
\label{firstpage}

\begin{abstract}
In an attempt to determine the outer scale of turbulence driven by
localized sources, such as supernova explosions in the interstellar
medium, we consider a forcing function given by the gradient of gaussian
profiles localized at random positions.
Different coherence times of the forcing function are considered.
In order to isolate the effects specific to the nature of the forcing
function we consider the case of a
polytropic equation of state and restrict ourselves to forcing amplitudes
such that the flow remains subsonic.
When the coherence time is short, the outer scale agrees with the half-width
of the gaussian.
Longer coherence times can cause extra power at large scales, but this
would not yield power law behavior at scales larger than that of the
expansion waves.
At scales smaller than the scale of the expansion waves the spectrum is
close to power law with a spectral exponent of $-2$.
The resulting flow is virtually free of vorticity.
Viscous driving of vorticity turns out to be weak and self-amplification
through the nonlinear term is found to be insignificant.
No evidence for small scale dynamo action is found in cases where
the magnetic induction equation is solved simultaneously with the other 
equations.
\end{abstract}


\section{Introduction}

Turbulence plays an important role in many branches of astrophysics.
During the past decade simulations have been employed
to address fundamental questions in turbulent stellar convection
zones \citep{Spruit_etal90}, accretion discs \citep{BalbusHawley98},
and interstellar turbulence \citep{Korpi_etal99,Balsara_etal04,Dib}.
In these three cases the nature of energy injection is quite
different; convective instability, magneto-rotational instability,
and explicit driving through supernova explosions, respectively.
Nevertheless, important insights into the nature of astrophysical turbulence
have been obtained by studying turbulence forced on large scales by
adding random time-dependent large scale perturbations to the velocity.
This allows the study of the turbulent cascade of kinetic energy
from large scales to smaller scales, and eventually down to the
dissipative scale \citep[see, e.g.,][for a review]{BS05}.
While the scale of the aforementioned instabilities can indeed
be large (comparable with the system size), this is not so
evident in the case of random forcing by supernova explosions.
The energy released by supernova explosions, and to a lesser extend also
by stellar winds and outflows, suffices to power interstellar turbulence
\citep{Vazq96}.
The anticipated outer scale of the turbulence is around 100\,pc 
\citep[e.g.,][]{BeckEtal96}.
This is large compared with the size of individual supernova explosion sites.
Some supernova remnants can stay reasonably coherent for scales
up to several tens of parsec.
Furthermore, supernova explosions are known to be clustered,
forming thereby large expanding patches called superbubbles
\citep[e.g.,][]{NI89,Fer92}.
Their scale is often larger than the scaleheight of the galactic disc,
and they are primarily responsible for driving fountain-like outflows.
In any case, it appears that the localized point-like supernova
explosions are able to act as a driver of the turbulence on a much
larger scale.

An obvious question concerns the connection between the artificial
forcings assumed in many computer simulations
and any of the more realistic types of forcing.
In the present paper we consider the properties of turbulence driven
by random, localized expansion waves, mimicking certain aspects of
supernova-driven explosions.
One of the questions we are able to address with such a setup
concerns information about the scale of the energy-carrying eddies
of turbulence that is driven by small localized expansion waves.
Is it true that such localized energy injections produce
kinetic energy predominantly at the scale of the original expansion
wave, even though this scale is in general quite small?
What are the effects, if any, that could shift the dominant power
to larger scales.
Large scales might plausibly arise if the duration of energy injection
at one particular site becomes comparable to or larger than the
turnover time of the resulting turbulence.
This would reinforce the original expansion wave until it has grown
to a larger size.
These issues are not directly connected with compressibility or with
the effects of cooling or the nature of the equation of state of the gas.
Thus, although interstellar turbulence is certainly highly supersonic,
we consider here the case of subsonic turbulence in order to make 
the connection between solenoidal forcing in the form of plane waves
and irrotational forcing driving localized expansion waves.

\section{Method}

We solve the compressible Navier-Stokes equations assuming a polytropic
equation of state relating the pressure $p$ to the density $\rho$ via
$p=K\rho^\gamma$ with $\gamma=5/3$ and polytropic coefficient
$K=c_{\rm s0}^2/(\gamma\rho_0^{\gamma-1})$, where
$c_{\rm s0}$ and $\rho_0$ are constants.
The momentum equation can then be written in the form
\EQ
\frac{\DD \uu}{\DD t}=-\nab h +\ff+\FF_{\rm visc},
\label{dudt}
\EN
where
\EQ
h={\rho_0 c_{\rm s0}^2\over\gamma-1}
\left({\rho\over\rho_0}\right)^{\gamma-1}
\EN
is the enthalpy,
$\DD/\DD t=\partial/\partial t+\uu\cdot\nab$ is the advective
derivative, and
$\FF_{\rm visc}=\rho^{-1}\nab\cdot\left(2\rho\nu\SSSS\right)$ is
the viscous force, where
${\sf S}_{ij}=\frac{1}{2}(u_{i,j}+u_{j,i})
-\frac{1}{3} \delta_{ij}\nab\cdot\uu$
is the traceless rate of strain tensor.
The density obeys the continuity equation,
\begin{equation}
\frac{\DD\ln\rho}{\DD t}=-\nab\cdot\uu.
\end{equation}
We adopt a gaussian potential forcing function $\ff$ of the form
\EQ
  \ff(\xx,t)=\nab\phi
\EN
with
\EQ
  \phi(\xx,t)=N\exp\left\{[\xx-\xx_{\rm f}(t)]^2/R^2\right\},
\EN
where $\xx=(x,y,z)$ is the position vector,
$\xx_{\rm f}(t)$ is the random forcing position,
$R$ is the radius of the gaussian, and $N$ is a normalization factor.
We consider two forms for the time dependence of $\xx_{\rm f}$.
First, we take $\xx_{\rm f}$ such that the forcing is 
$\delta$-correlated in time. Second, we include a forcing time 
${\delta{}t}_{\rm force}$ that defines the interval during which $\xx_{\rm f}$ 
remains constant. On dimensional grounds the normalization is chosen to be
$N=c_{\rm s0}\sqrt{c_{\rm s0}R/\Delta{}t}$, where
\EQ
\Delta{}t=\max\left(\delta{}t,\delta{}t_{\rm force}\right)
\EN
is the length of the time step, $\delta t$,
in the $\delta$-correlated case or equal to the mean
interval $\delta{}t_{\rm force}$ during which the force remains unchanged,
depending on which is longer.
We begin by considering the nature of flows generated in each case in the 
absence of a magnetic field.
We use the \textsc{Pencil Code},\footnote{
\url{http://www.nordita.dk/software/pencil-code}} which is a non-conservative,
high-order, finite-difference code (sixth order in space and
third order in time) for solving the compressible hydrodynamic and
hydromagnetic equations.
We adopt non-dimensional variables by measuring
speed in units of the sound speed, $c_{\rm s0}$, and length in units of
$1/k_1$, where $k_1$ is the smallest wavenumber in the periodic domain.
This implies that the nondimensional size of the domain is $(2\pi)^3$.

\section{Results}

\subsection{Delta-correlated forcing function}

\begin{figure}\centering
\includegraphics[width=\columnwidth]{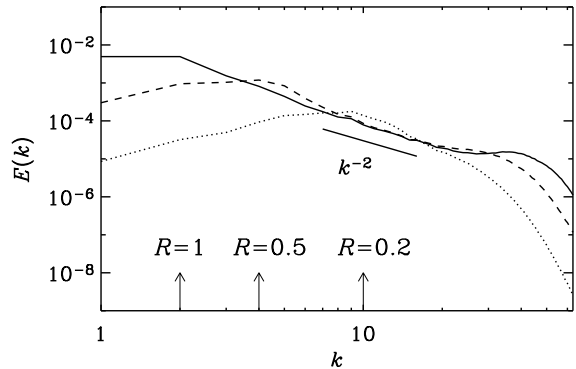}
\caption{
Time averaged energy spectra for $R=0.2$ (dotted line),
0.5 (dashed line), and 1 (solid line).
The durations of the runs are $N_{\rm turn}=333$, 224, and 81
turnover times, respectively.
$128^3$ mesh points, $\nu=10^{-3}$, and $\delta$-correlated forcing.
The vertical arrows indicate the effective forcing wavenumber,
as obtained from \Eq{kpeak}.
}\label{pcomp_spec1}\end{figure}

We begin with the case where $\delta t_{\rm force}=0$, so the forcing
function is $\delta$-correlated in time.
In \Figs{pcomp_spec1}{pcomp_spec2} we show the resulting time averaged
energy spectra for two different resolutions.
The results are relatively robust with respect to changes in the
resolution.
Also, the spectra are found to converge quite quickly after only about
5--7 turnover times, suggesting that there are no strong transients.
For the results presented below, the durations of the runs, $T_{\rm run}$,
is given in terms of the turnover time,
\EQ
N_{\rm turn}=u_{\rm rms} k_{\rm peak} T_{\rm run},
\EN
where $k_{\rm peak}$ is the effective forcing wavenumber which, in turn,
depends on the radius of the initial expansion wave, $R$.
Indeed, the energy spectra show maximum power at the wavenumber
$k_{\rm peak}$, which is approximately inversely proportional to $R$, with
\EQ
k_{\rm peak}R\approx2.
\label{kpeak}
\EN
The wavenumber $k_{\rm peak}$ agrees with the width of
the Fourier-transformed forcing function.
[Note that the Fourier transform of $\exp(-x^2/R^2)$ is $\exp(-k^2R^2/4)$.]
This result is therefore not surprising, but it confirms quite clearly
that a $\delta$-correlated forcing function consisting of localized
expansion waves produces maximum energy injections at the scale
of the expansion wave itself.
This would therefore {\it not} explain the
energy spectrum seen for example in simulations of supersonic
turbulence \citep{Porter98,Padoan99,Haugen04} or those
inferred for interstellar turbulence \citep{Han_etal04}, where energy
injection is assumed to occur at the scale of the domain.

\begin{figure}\centering
\includegraphics[width=\columnwidth]{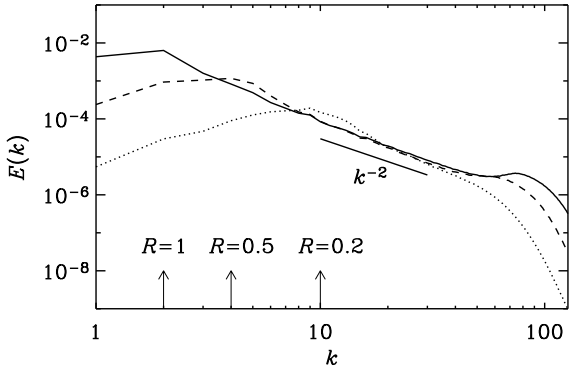}
\caption{
Same as \Fig{pcomp_spec1}, but for $256^3$ mesh points, $\nu=5\times10^{-4}$,
for $R=0.2$ (dotted line), 0.5 (dashed line), and 1 (solid line).
The durations of the runs are $N_{\rm turn}=120$, 64, and 29
turnover times, respectively.
}\label{pcomp_spec2}\end{figure}

The energy spectrum shows a distinct $k^{-2}$ subrange.
Such a slope is predicted for shock turbulence \citep{KP73},
although here the flow is subsonic and without shocks.
However, it may be interesting to note that a $k^{-2}$ spectrum
has also been seen in the irrotational component of transonic
turbulence \citep{Porter98}.
Energe\-tically, the irrotational component is subdominant compared
with the solenoidal components \citep{Porter98,Padoan99,Haugen04}.
In the present simulations, however, there is hardly any vorticity
and the flow is entirely dominated by the irrotational component.
This, in turn, is a consequence of the irrotational nature of the
forcing function combined with the fact that the viscosity is low,
so that very little vorticity is produced.
We consider vorticity production in more detail in
\Sec{VorticityProduction} below.

\subsection{Forcing function with memory}

\begin{figure}\centering
\includegraphics[width=\columnwidth]{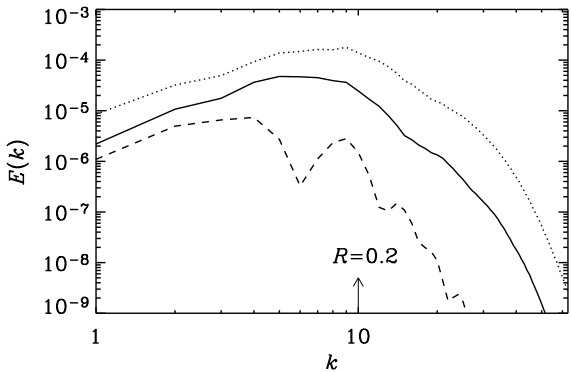}
\caption{
Time averaged energy spectra for $R=0.2$ for
$\delta t_{\rm force}=1$ ($\mbox{St}=0.08$, dashed line),
$\delta t_{\rm force}=0.2$ ($\mbox{St}=0.05$, solid line), and
$\delta t_{\rm force}=0$ (i.e.\ $\delta$-correlated, $\mbox{St}=0$, dotted line).
The durations of the runs are $N_{\rm turn}=16$, 64, and 333
turnover times, respectively, and the rms velocities are
0.008, 0.027, and 0.058, respectively.
$128^3$ mesh points, $\nu=10^{-3}$.
}\label{pcomp_spec1dt}\end{figure}

Next we look at the case where the expansion wave persists for
a certain amount of time, $\delta t_{\rm force}$.
In \Fig{pcomp_spec1dt} we show time averaged spectra for different values
of $\delta t_{\rm force}$.
This time interval is conveniently expressed in non-dimensional form
as a Strouhal number \citep{LL87,KR80},
\EQ
\mbox{St}=\delta t_{\rm force}u_{\rm rms}k_{\rm peak}.
\EN
Note that for larger values of $\delta t_{\rm force}$
the spectrum consists of separated bumps, but the overall
power appears to be shifted to larger scales and smaller
wavenumbers.
The presence of bumps is suggestive of an acoustic resonance or standing
wave pattern that may be excited in the simulation box.
Such a phenomenon is unfamiliar in the context of mostly vortical
turbulence.
It is plausible, and indeed compatible with our as yet limited results, that
these bumps can only occur if there is sufficient scale separation
between $k_{\rm peak}$ and $k_1$.
Furthermore, bumps in the spectrum seem to be possible only when
the memory time is not too short, i.e.\ when the effects of temporal
randomness are limited.

\begin{figure}\centering
\includegraphics[width=\columnwidth]{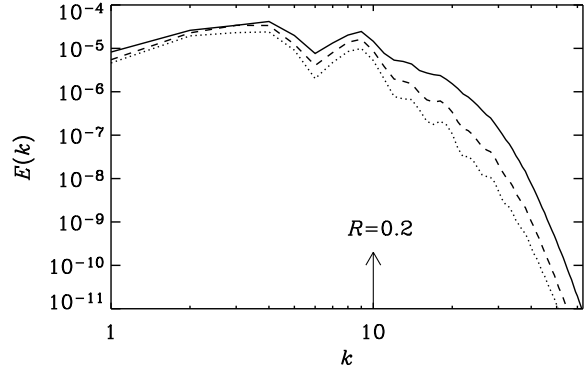}
\caption{
Dependence on viscosity.
Time averaged energy spectra for 
$\nu=2\times10^{-4}$ (solid line),
$\nu=5\times10^{-4}$ (dashed line),
$\nu=10^{-3}$ (dotted line).
The durations of the runs are $N_{\rm turn}=28$, 17, and 138
turnover times, respectively, and the values of the St are
0.23, 0.19, and 0.17, respectively.
$\delta t_{\rm force}=1$, $R=0.2$, $128^3$ mesh points.
}\label{pcomp_spec1dt_nu}\end{figure}

In \Figs{pcomp_spec1dt_nu}{pcomp_spec1dt_res} we demonstrate that
decreasing viscosity and increasing resolution produces more power at
small scales, gradually building up a $k^{-2}$ power law between the
forcing scale and the dissipative cutoff scale when the  
resolution is sufficient for those scales to be resolved.

\begin{figure*}\centering
\includegraphics[width=.246\textwidth]{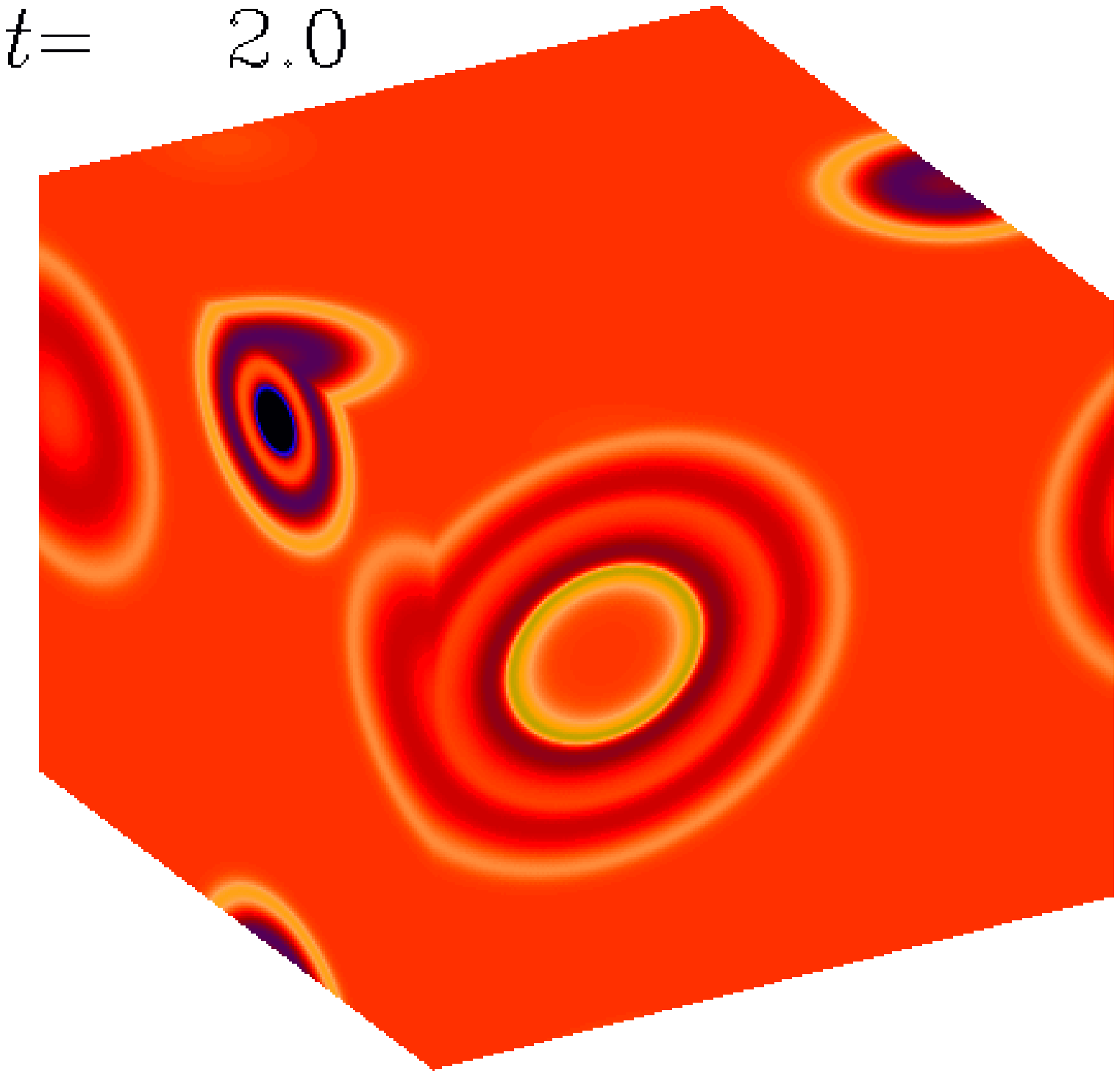}
\includegraphics[width=.246\textwidth]{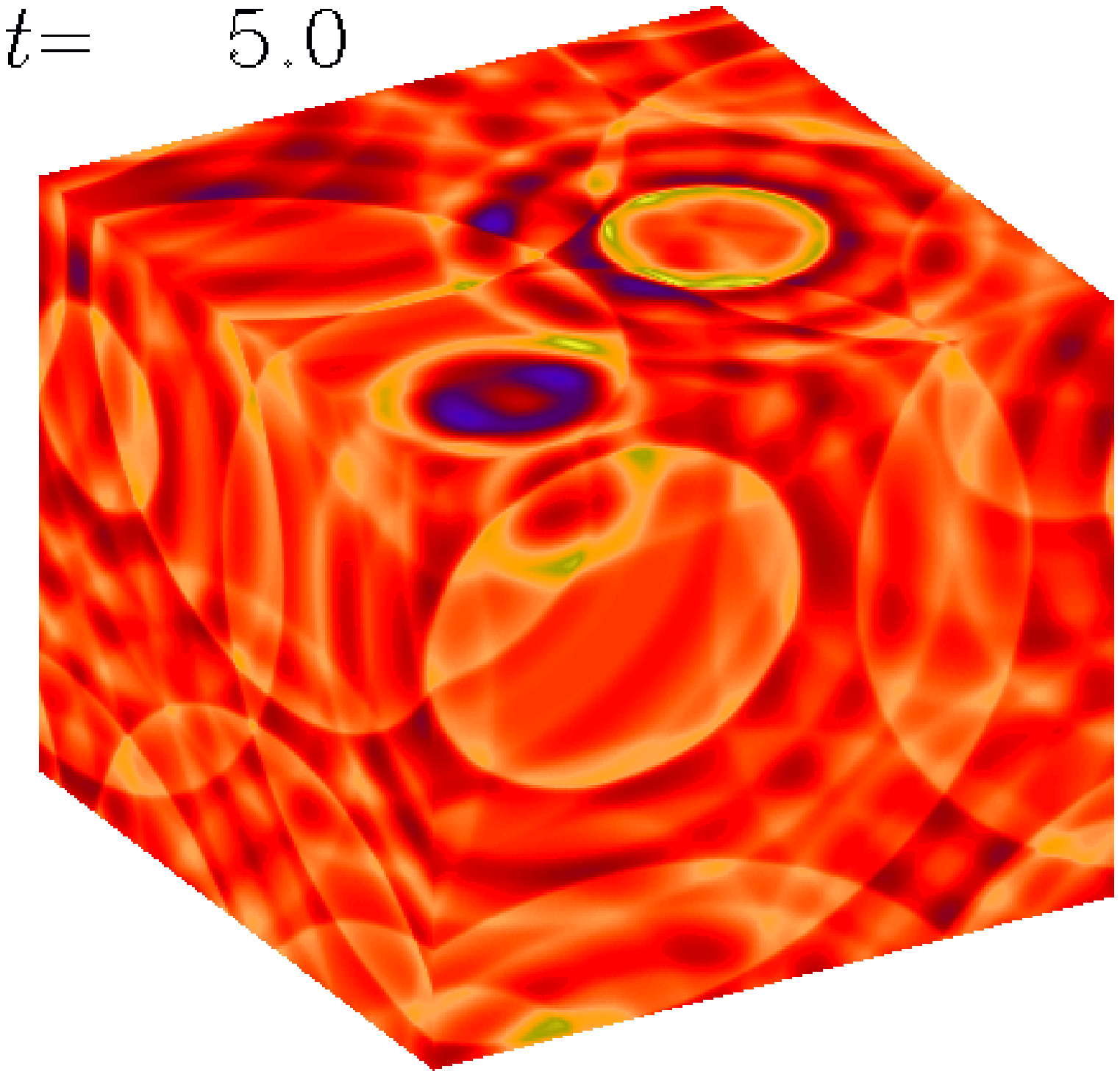}
\includegraphics[width=.246\textwidth]{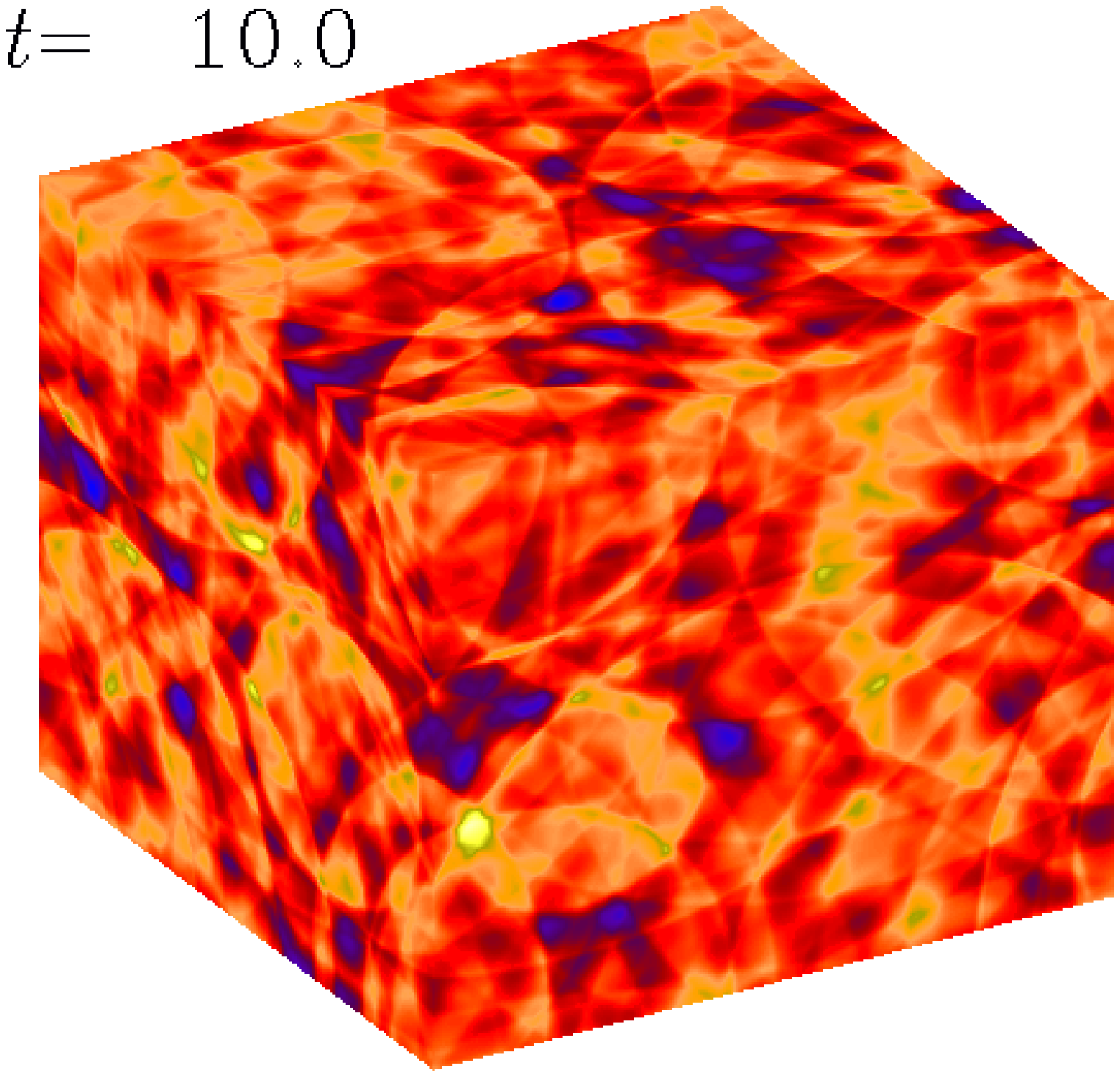}
\includegraphics[width=.246\textwidth]{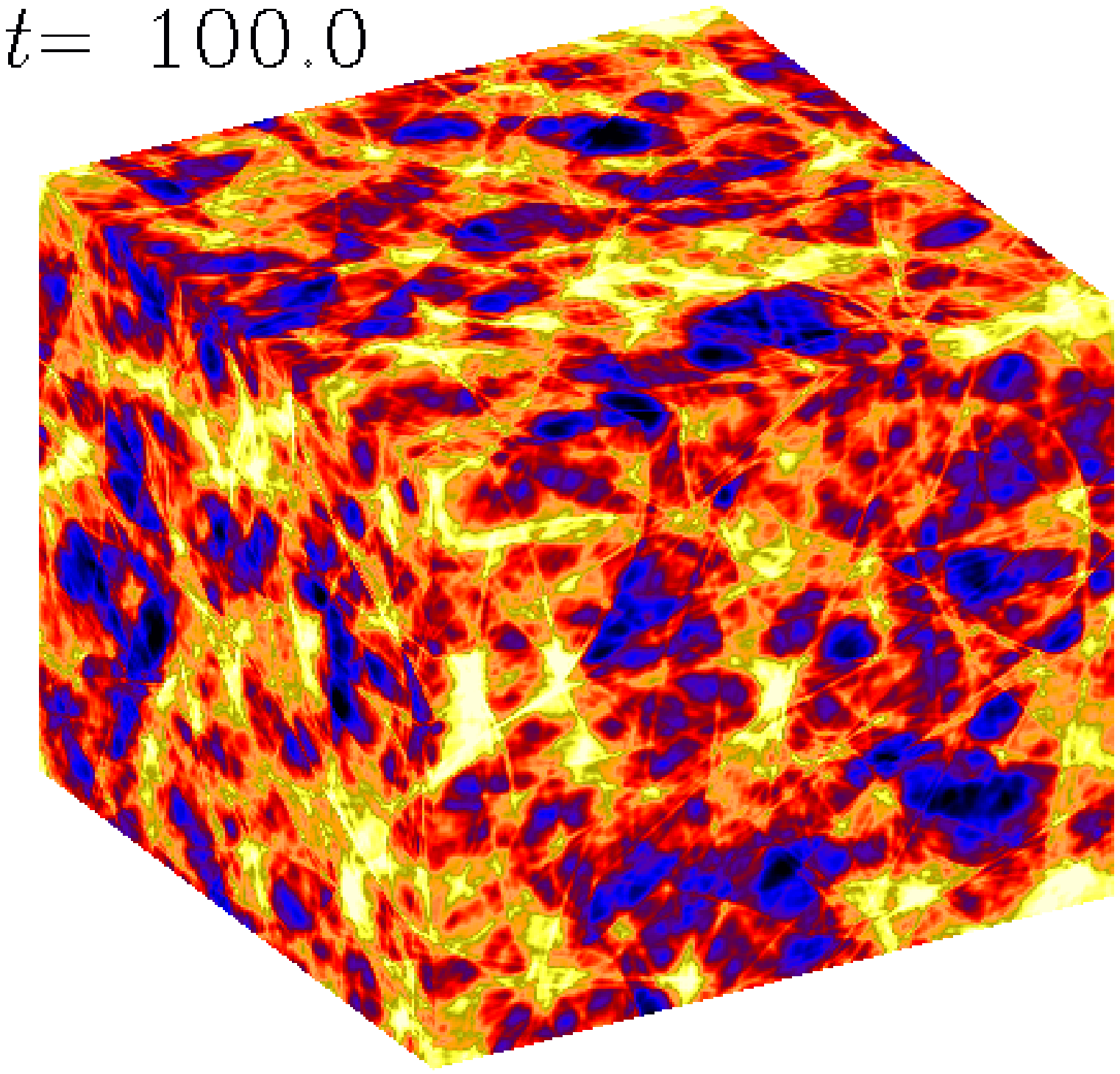}
\caption{
Visualization of $\ln\rho$ on the periphery of the box at different times,
for $k_1 R=0.2$,$\mbox{Re}=50$, $\mbox{St}=0.25$, $\nu=5\times10^{-5}$,
and $512^3$ mesh points.
Note that in the fully developed state individual expansion waves
can hardly be recognized.
}\label{img}\end{figure*}

Visualizations of the density (\Fig{img}) show how the initially highly
ordered expansion waves turn rapidly into a complicated
pattern.\footnote{Animations of the density can be found at
\url{http://www.nordita.dk/~brandenb/movies/gaussianpot}}
By the time $t=50$ the velocity has reached a statistically 
steady state and the flow has ceased to bear any resemblance to
the initial expansion wave.

\begin{figure}\centering
\includegraphics[width=\columnwidth]{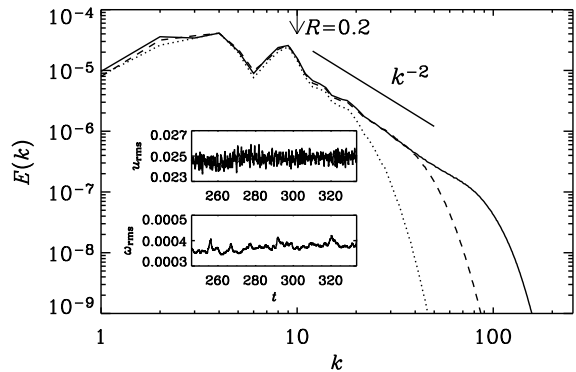}
\caption{
Dependence on resolution.
Time averaged energy spectra for $R=0.2$,
using $512^3$ mesh points (solid line, $\nu=5\times10^{-5}$),
$256^3$ mesh points (dashed line, $\nu=10^{-4}$), and
$128^3$ mesh points (dotted line, $\nu=2\times10^{-4}$).
The durations of the runs are $N_{\rm turn}=22$, 59, and 28
turnover times, respectively,
and the values of St are 0.25, 0.24, and 0.23, respectively.
The insets shows the evolution of $u_{\rm rms}$ and $\omega_{\rm rms}$
for the run with $512^3$ mesh points.
}\label{pcomp_spec1dt_res}\end{figure}

\subsection{Vorticity production}
\label{VorticityProduction}

One of the main differences between the present forcing function and
those used in some previous papers \citep{EP88,Porter98,Padoan02,Haugen04}
is that here the forcing is completely irrotational,
i.e.\ $\nab\times\ff=0$ and $\nab\cdot\ff\neq0$.
Therefore, unlike previous cases where instead $\nab\times\ff\neq0$
and $\nab\cdot\ff=0$, and where vorticity ($\oo=\nab\times\uu$) was
immediately produced, here vorticity can only be produced through
viscous interactions.
This can be seen by writing the viscous force in the form
\EQ
\FF_{\rm visc}=-\nu\nabla\times\nabla\times\uu
+{\textstyle{4\over3}}\nu\nab\nab\cdot\uu+\nu\GG,
\EN
where $G_i={\sf S}_{ij}\nabla_j\ln\rho$ is a term that could drive
vorticity even with zero vorticity initially.
Indeed, the vorticity equation, obtained by taking the curl of \Eq{dudt},
can then be written as
\EQ
{\partial\oo\over\partial t}=\nab\times(\uu\times\oo)+\nu\nabla^2\oo
+\nu\nab\times\GG.
\label{dodt}
\EN
Note the absence of the baroclinic term, i.e.\ the cross product of
temperature and specific entropy gradients.
This is because we have restricted ourselves to a gas with an isothermal
equation of state.
The baroclinic term is known to produce vorticity \citep[e.g.][]{Korpi_etal99b}.
The nonlinear term, $\nab\times(\uu\times\oo)$, could in principle lead
to exponential amplification of an initial seed vorticity---much like a term
in the small scale dynamo problem, discussed in the next section.
However, no evidence for such an effect has been found in the present
simulations.
Thus, the only remaining term is $\nab\times\GG$ which is due to viscosity.
One might therefore expect that as we decrease the Reynolds number,
defined here in terms of $k_{\rm peak}$ as
\EQ
\mbox{Re}=u_{\rm rms}/\left(\nu k_{\rm peak}\right),
\EN
the amount of vorticity production decreases.
The numerical simulations yield values for the normalized vorticity,
\EQ
\mbox{normalized vorticity}=\omega_{\rm rms}/(u_{\rm rms}k_{\rm peak}),
\EN
that are essentially within the `noise'.
Indeed, the data shown in \Tab{Tsum} are well below unity and do not give a
clear trend as a function of resolution or Reynolds number.
For all these runs the Strouhal number is $\mbox{St}\approx0.25$.

\begin{table}\caption{
Results for the normalized vorticity,
$\omega_{\rm rms}/(u_{\rm rms}k_{\rm peak})$,
as a function of Re and resolution.
The durations of the runs vary between $N_{\rm turn}=16$ and 250
turnover times, and $\nu$ varies between $10^{-3}$ and $5\times10^{-5}$.
For the high resolution run with $512^3$ meshpoints and
$\nu=5\times10^{-5}$ we have $N_{\rm turn}=59$ turnover times.
}\vspace{12pt}\centerline{\begin{tabular}{cccc}
Re & $512^3$ & $256^3$ & $128^3$ \\
\hline
50 & $1.4\times10^{-3}$ & $8.7\times10^{-3}$ &  \\
25 &                    & $1.1\times10^{-2}$ & $2.9\times10^{-3}$ \\
12 &                    & $1.6\times10^{-2}$ & $2.0\times10^{-3}$ \\
 4 &                    & $7.6\times10^{-3}$ & $1.5\times10^{-3}$ \\
\label{Tsum}\end{tabular}}\end{table}


In conclusion, the amount of vorticity production depends
mostly on the numerical resolution, suggesting that no
measurable vorticity is produced by physical effects.

\subsection{Dynamo action?}

In view of the formal analogy between the vorticity and induction
equations \citep{Bat50}, it appears that dynamo action should be
as difficult to achieve as the amplification of vorticity by the
nonlinear term, $\nab\times(\uu\times\oo)$.
In order to address this problem, we solve the induction equation
for the magnetic field $\BB$,
\EQ
{\partial\BB\over\partial t}=\nab\times(\uu\times\BB)+\eta\nabla^2\BB,
\label{dBdt}
\EN
simultaneously with the other equations.
\EEq{dudt} gains a term---the Lorentz force per unit mass,
$\JJ\times\BB/\rho$, but this effect would only become important
once the magnetic field becomes strong.
(Here, $\JJ=\nab\times\BB/\mu_0$ is the current density and $\mu_0$
is the vacuum permeability.)
As initial condition we adopt a magnetic vector potential that is
$\delta$-correlated in space.
This results in a magnetic energy spectrum proportional to $k^4$.

\begin{figure}\centering
\includegraphics[width=\columnwidth]{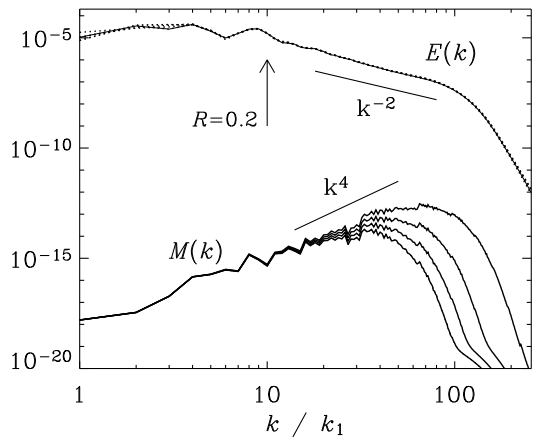}
\caption{
Kinetic and magnetic energy spectra, $E(k)$ and $M(k)$, respectively,
at three different times, separated by $\Delta t=20\delta t_{\rm force}$,
with for $R=0.2$, $R_{\rm m}=u_{\rm rms}/(\eta k_{\rm peak})=250$,
$\mbox{Re}=50$, and $512^3$ mesh points.
The duration of the run is $N_{\rm turn}=22$ turnover times.
$\nu=5\times10^{-5}$, $\eta=5\times10^{-5}$.
}\label{pspec512b3}\end{figure}

In all the cases that we have investigated so far, we have not found
evidence of field amplification, i.e.\ there is no dynamo action.
Our highest resolution run has $512^3$ mesh points and a magnetic
Reynolds number, $R_{\rm m}=u_{\rm rms}/(\eta k_{\rm peak})$,
of 250 and $\mbox{Re}=50$.
Kinetic and magnetic energy spectra are shown in \Fig{pspec512b3}.
For this run the magnetic decay rate, $\lambda=-\dd\ln B_{\rm rms}/\dd t$,
is about 30 times the ohmic decay rate, $\eta k_{\rm peak}^2$.
This is approximately equal to the scale separation ratio between
$k_{\rm peak}$ and the dissipative cutoff wavenumber.
Such a rapid decay rate suggests that this type of flow has a highly destructive
effect on the magnetic field.
Energy spectra of the magnetic field confirm that the spectral energy decays
at small scales, while it stays unchanged at large scales.

The absence of any dynamo action agrees with earlier results
for purely irrotational turbulence \citep{Brooks99}.
On the other hand, earlier analytic considerations suggested that
dynamo action should be possible for irrotational flows and that
the growth rate should increase with increasing Mach number
to the fourth power \citep{KRS85,MS96}.
Subsequent work by \cite{Haugen04} did not discuss growth rates,
but critical magnetic Reynolds numbers which shows that
the small scale (non-helical) dynamo
becomes about twice as hard to excite when the Mach number exceeds unity.
This was explained by the presence of an additional irrotational
contribution in the supersonic case.
This contribution remained however subdominant and did not contribute
to the dynamo.
This is consistent with the present result that up to the largest magnetic
Reynolds numbers accessible today, purely irrotational turbulence does
not produce dynamo action.

\section{Conclusion}

The localized expansion waves used as forcing functions in the present
paper resemble in some ways the driving by supernova explosions.
An obvious difference is that the supernova driving produces highly
supersonic flows.
This may be important in connection with understanding the amount of
vorticity production in irrotationally forced flows.
Indeed, \cite{Korpi_etal99b} found that the main driver of vorticity
production in their supernova driven turbulence simulations is the
baroclinic term.

The present work suggests that the turbulence would indeed be peaked
primarily at small scales.
Memory effects (i.e.\ finite values of ${\delta{}t}_{\rm force}$)
contribute somewhat to enhancing power at larger scales,
but, because of the dip in power at intermediate scales, the resulting
energy spectrum is still very different from the power law spectra
obtained with forcing directly at large scales.
The main difference is the absence of vorticity in the present simulations.
Since vorticity is mainly driven by the baroclinic term, we may expect
that it is also this term that would primarily contribute to enhanced
power at low wave numbers, corresponding to scales larger than the
radius of the original expansion waves. 

\section*{Acknowledgments}
We thank Maarit Korpi for discussions and useful suggestions on an
earlier draft of the manuscript, and an anonymous referee for detailed
comments that have helped to improve the presentation.
This work was supported by PPARC grant {PPA/S/S/2002/03473}.
AJM is grateful to Nordita for financial support and hospitality.
We acknowledge the Danish Center for Scientific Computing
for granting time on the Horseshoe cluster.


\label{lastpage}
\end{document}